# Channel Reconstruction for SVD-ZF Precoding in Massive 3D-MIMO Systems: Low-Complexity Algorithm

Yuwei Ren[1], Xin Su[2], Can Qi[3], Yingmin Wang[2]

[1] Beijing University of Posts and Telecommunications, Beijing, China
[2] State Key Laboratory of Wireless Mobile Communications, China Academy of Telecommunications Technology, Beijing, China
[3] Langfang Power Supply Company, Jibei Electric Power Company Limited, Langfang, China
Email: yuweir@gmail.com, wangyingmin@catt.cn

**Abstract**: In this paper, we study the low-complexity channel reconstruction methods for downlink precoding in massive MIMO systems. When the user is allocated less streams than the number of its antennas, the BS or user usually utilizes the singular value decomposition (SVD) factorizations to get the effective channels, whose dimension is equal to the num of streams. This process is called channel reconstruction in BS for TDD mode. However, with the increasing of antennas in BS, the computation burden of SVD is becoming incredibly high. As a countermeasure, we propose a series of novel low-complexity channel reconstruction methods for downlink zero-forcing precoding (ZF). We adopt randomized algorithms to construct an approximate SVD, which could reduce the dimensions of the matrix, especially when approximating an input matrix with a low-rank element. Besides, this method could automatically modify the parameters to adapt arbitrary number demand of streams from users. The simulation results show that the proposed methods only cost less than 30% float computation than the traditional SVD-ZF method, while keeping nearly the same performance of 1Gbps with 128 BS antennas.
**Key words**: Channel Reconstruction; SVD; 3D-MIMO; Massive MIMO;

## I. INTRODUCTION

Massive MIMO is one of the most importantly investigated subjects in the literature of coming 5G technologies due to the high potential it offers in improving not only the reliability but also the throughput of the system. Information theory has shown that the optimum capacity of MU-MIMO channels could be achieved through simple Zero-forcing (ZF) methods in large scale arrays [1].

But subject to the actual physical space, the number of antennas aiding the BS cannot go to infinity. The use of active antenna systems (AAS) and uniform planar array (UPA) draw much attention and make practical massive MIMO possible. Some recent deployments have shown up to 30 percent gain in system capacity by taking advantage of elevation domain beams (e.g., [2, 3]) in UPA system.

However, the deployment in realistic setup is hindered by several practical challenges that are not of concern in conventional MIMO systems. The computational complexity becomes a main concern when the number of antenna and user goes to be large or infinite. For example, the burden of matrix decomposition in channel reconstruction would be a big challenge. The traditional singular value decomposition (SVD) usually requires $O(Nt^2)$ float computations [4], where $Nt$ is the number of BS antennas. With $Nt$ increasing, this is an inconceivable challenge for hardware implementation.

Much work has been completed in the face of such challenge. For example, [5] [6] give some iterative convergence algorithms to approximate matrix SVD, which is widely used in traditional MIMO systems. But with the BS antennas increasing, the convergence would not be guaranteed in limited iteration, which may lead to large matrix computation. What's more, 3GPP is actively developing the 3D channel model to enable the elevation beamforming. For this, [7][8] apply the 3D beamforming to massive MIMO, where the elevation and azimuth antennas make beamforming respectively, and the two beamforming vectors are transformed into precoding matrix by Kronecker product. But this method is limited by the special antennas structure with elevation array. Besides, [4] have summarized many fast and simplified methods for direct SVD computation, (e.g., Hestenes-Jacobi matrix theory and its extension). But Jacobi's method is based on general setting for SVD and don't fully utilize the special constraints in communication, e.g., streams constraints and correlation constraints among antennas. So the corresponding complexity is still very high.

In this paper, we address the challenge of computation complexity caused by matrix SVD, and propose a series of new channel reconstruction methods based on randomized algorithms. From the simulation, the proposed methods could greatly reduce the complexity while keeping acceptable performance. The novelty mainly includes three parts:
➢ Firstly, our methods utilize a random small-size matrix to construct the equivalent range of the target large matrix. And the matrix factorization in small equivalent range could approximate

- large dimensional SVD with less complexity.
- Secondly, the methods directly eliminate the redundancy from the difference between the number of UE antennas and effective streams. It only considers the equivalent dimension matrix SVD with effective streams. Compared to some direct fast SVD methods based on Jacobi [4], the additional burden introduced by the number of UE antennas is greatly reduced by low dimension SVD.
- Thirdly, our methods could cope with the demands of different number of user's streams. These parameters in these methods could be freely changed according to the realistic setting, with the constraints, such as the number of effective channel rank and BS/user's antennas, capacity performance and complexity.

The 3D channel model under 3D-UMi scenario calibrated by 3GPP [9] is adopted. In MU-MIMO system, the link-level performance of the proposed SVD-ZF methods and the complexity are analyzed and compared. The rest of this paper is organized as follows. Section II describes the problem formulation. The proposed low-complexity channel reconstruction methods are presented in Section III. Besides, it outlines complexity analysis. The simulation is performed in Section V. And the last Section gives the conclusion.

Notation: Bold letters are utilized to denote the matrix or the vector. $\mathbf{I}_K$ is the identity matrix of $K \times K$. $CN(\mathbf{m}, \mathbf{R})$ denotes the circular symmetric complex Gaussian distribution with mean $\mathbf{m}$ and covariance matrix $\mathbf{R}$. The superscripts T, H denote the transpose and conjugate transpose, respectively. The Kronecker product is denoted by $\otimes$.

## II. SYSTEM MODEL AND PROBLEM FORMULATION

### 2.1 3D channel model

Relative to the traditional 2D SCM channel model, the 3D channel model considers the radio propagation in the elevation dimension. Here the 3D channel model calibrated by the 3GPP is recommended [9].

The generation of the large scale parameters can be referred to [9]. The small-scale channel coefficient is generated by summing the contribution of some rays. Assume that one ray is composed of $M_L$ sub-paths. The coordinate system for 3D channel model can be seen from Fig.1. $ray_{m,n}$ means the m-th sub-path in the n-th ray. The global coordinate system defines the zenith angle $\theta_{ZoD/ZoA}$ and the azimuth angle $\phi_{AoD/AoA}$. $\theta_{ZoD/ZoA} = 90^0$ points to the horizontal direction and $\theta_{ZoD} = 0°$ points to the zenith direction[9]. $\hat{n}$ is the given direction of $ray_{m,n}$. $\hat{\varphi}$ and $\hat{\theta}$ are the spherical basis vectors.

The channel coefficient from transmitter element s to receiver element u for the n-th ray is modeled as:

For NLOS path,

$$H_{u,s,n}(t) = \sqrt{P_n/M} \sum_{m=1}^{M} \begin{bmatrix} F_{rx,u,\theta}(\theta_{n,m,ZOA}, \varphi_{n,m,AOA}) \\ F_{rx,u,\varphi}(\theta_{n,m,ZOA}, \varphi_{n,m,AOA}) \end{bmatrix}^T \begin{bmatrix} \exp(j\Phi_{n,m}^{\theta\theta}) & \sqrt{\kappa_{n,m}^{-1}} \exp(j\Phi_{n,m}^{\theta\varphi}) \\ \sqrt{\kappa_{n,m}^{-1}} \exp(j\Phi_{n,m}^{\varphi\theta}) & \exp(j\Phi_{n,m}^{\varphi\varphi}) \end{bmatrix}$$

$$\begin{bmatrix} F_{tx,s,\theta}(\theta_{n,m,ZOD}, \varphi_{n,m,AOD}) \\ F_{tx,s,\varphi}(\theta_{n,m,ZOD}, \varphi_{n,m,AOD}) \end{bmatrix} \exp\left(j2\pi\lambda_0^{-1}\left(\hat{r}_{rx,n,m}^T \cdot \bar{d}_{rx,u}\right)\right) \exp\left(j2\pi\lambda_0^{-1}\left(\hat{r}_{tx,n,m}^T \cdot \bar{d}_{tx,s}\right)\right) \exp(j2\pi v_{n,m} t)$$

For LOS path,

$$H_{u,s,n}(t) = \sqrt{\frac{1}{K_R+1}} H_{u,s,n}^{'}(t)$$

$$+ \delta(n-1) \sqrt{\frac{K_R}{K_R+1}} \begin{bmatrix} F_{rx,u,\theta}(\theta_{LOS,ZOA}, \varphi_{LOS,AOA}) \\ F_{rx,u,\varphi}(\theta_{LOS,ZOA}, \varphi_{LOS,AOA}) \end{bmatrix}^T \begin{bmatrix} \exp(j\Phi_{LOS}) & 0 \\ 0 & -\exp(j\Phi_{LOS}) \end{bmatrix}$$

$$\begin{bmatrix} F_{tx,s,\theta}(\theta_{LOS,ZOD}, \varphi_{LOS,AOD}) \\ F_{tx,s,\varphi}(\theta_{LOS,ZOD}, \varphi_{LOS,AOD}) \end{bmatrix} \cdot \exp\left(j2\pi\lambda_0^{-1}\left(\hat{r}_{rx,LOS}^T \cdot \bar{d}_{rx,u}\right)\right) \cdot \exp\left(j2\pi\lambda_0^{-1}\left(\hat{r}_{tx,LOS}^T \cdot \bar{d}_{tx,s}\right)\right) \cdot \exp(j2\pi v_{LOS} t)$$

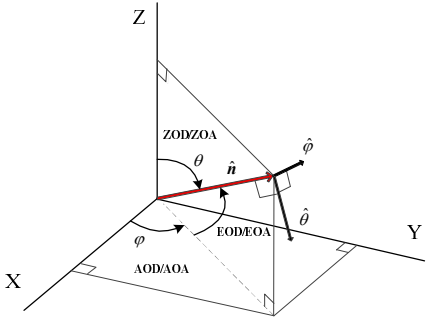

**Fig.1 The coordinate system for 3D channel model**

where $F_{rx,u,\theta}$ and $F_{rx,u,\phi}$ are the antenna radiation patterns for element u in the direction of the spherical basis vectors, $\hat{\theta}$ and $\hat{\phi}$ respectively. $F_{tx,s,\theta}$ and $F_{tx,s,\phi}$ are the antenna radiation patterns for element s in the direction of the spherical basis vectors, $\hat{\theta}$ and $\hat{\phi}$ respectively.

$$\hat{r}_{rx,n,m} = \begin{bmatrix} \sin\theta_{n,m,ZOA}\cos\varphi_{n,m,AOA} \\ \sin\theta_{n,m,ZOA}\sin\varphi_{n,m,AOA} \\ \cos\theta_{n,m,ZOA} \end{bmatrix}$$

is the unit vector about the azimuth of arrival angle(AoA)$\phi_{n,m,AOA}$ and the zenith of arrival angle(ZoA)$\theta_{n,m,ZOA}$. And also $\hat{r}_{tx,n,m}$ is the counterpart at the transmit side; $\bar{d}_{rx,u}$ and $\bar{d}_{tx,s}$ are the location vectors of the transmit and receive elements, respectively; $\{\Phi_{n,m}^{\theta\theta}, \Phi_{n,m}^{\theta\phi}, \Phi_{n,m}^{\phi\theta}, \Phi_{n,m}^{\phi\phi}\}$ are the random initial phases for sub-path m of ray n; $\lambda_0$ is the wavelength of the carrier frequency; $K_R$ is the Ricean K-factor; $v_{n,m}$ is the Doppler frequency component. More detailed description about the generation of 3D channel model can be referred to [9].

## 2.2 3D-MIMO system model

The BS in each cell is equipped with 2D planar cross-polarized antenna array. The number of antenna elements in azimuth direction and elevation direction is $N_A$ and $N_E$ respectively. And the total number of antennas is $Nt = N_A \times N_E \times 2$. The configuration of antenna array is presented in Fig.2. Each user is equipped with $M$ antennas.

The received signal $\mathbf{x}^k \in C^{S_k \times 1}$ of the k-th user in MU-MIMO can be expressed as:

$$\mathbf{x}^k = \sqrt{\rho_f}\mathbf{E}_{(S_k \times M)}^k \mathbf{H}_{(M \times Nt)}^k \mathbf{W}_{(Nt \times S_k)}^k \mathbf{s}_{(S_k \times 1)}^k$$
$$+ \sum_{l=1, l \neq k}^{K} \sqrt{\rho_f} \mathbf{E}^k \mathbf{H}^k \mathbf{W}^l \mathbf{s}^l + \mathbf{v}_k$$

where $M$, $Nt$, $S_k$ are the num of user antennas, BS antennas and the allocated streams for k-th user, respectively. $\sum_{l=1, l \neq k}^{K} \sqrt{\rho_f} \mathbf{E}^k \mathbf{H}^k \mathbf{W}^l \mathbf{s}^l$ represents the interferences from the other users. $\mathbf{H}_{(M \times Nt)}^k$ is the k-th downlink channel matrix, $K$ is the number of users, $\mathbf{W}_{(Nt \times S_k)}^k$ is the k-th precoding matrix. $\mathbf{E}_{(S_k \times M)}^k$ is the k-th estimating matrix. $\mathbf{s}_{(S_k \times 1)}^k$ is transmitted signal for the k-th user. $\mathbf{v}_k \sim CN(\mathbf{0}, \sigma^2 \mathbf{I}_{S_k \times S_k})$ is the noise. Here we consider ZF precoding method, and $\mathbf{W}_{(Nt \times S_k)}^k = \mathbf{H}_{(S \times Nt)}^H \left(\mathbf{H}_{(S \times Nt)}\mathbf{H}_{(S \times Nt)}^H\right)^{-1}$.

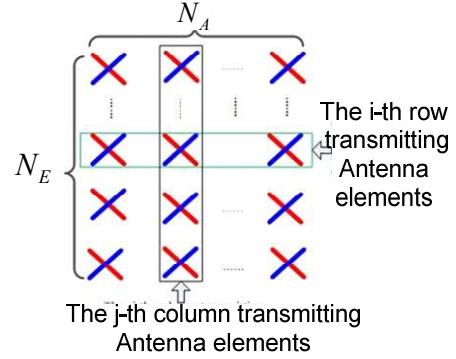

**Fig.2 The set of 2D planar cross-polarized antenna array, with $Nt = N_A \times N_E \times 2$ antennas and dual-polarized (0/+90).**

## 2.3 Problem formulation

In conventional LTE systems, the users are usually equipped with $M$ antennas (e.g., 2, 4, 8 or else), and allocated $S$ layers (e.g., 1, 2 or else), which means that BS has to extract $S*Nt$ effective channels from $M*Nt$ estimated user channels before downlink precoding. In fact, the number of effective data steams is usually less than of user's antennas ($M \geq S$) in order to maintain the high performance. The whole process is shown in Fig.3. In TDD-MIMO systems, the users send sounding reference signals (SRS) to the BS in the uplink, and then, the BS estimates these SRS and gets the channels $\mathbf{H}_{M \times Nt}$ for each user in step 1. But how does the BS get the effective user channels from $M*Nt$ channel matrix to $S$ streams? The measure is to make channel reconstruction by the SVD in $\mathbf{H}_{M \times Nt}$, which is to get $S$ main singular-vectors for $\mathbf{H}_{S \times Nt}$. Next, BS gets downlink effective channel $\mathbf{H}_{S \times Nt}$ by TDD reciprocity and makes the downlink precoding. Therefore, our focus is the Step: 2 ($\mathbf{H}_{M \times Nt} \rightarrow \mathbf{H}_{S \times Nt}$).

## III. SVD FACTORIZATION METHODS

### 3.1 Traditional channel reconstruction method

After estimating the SRS in TDD-LTE uplink, BS gets the realistic user's MIMO channel $\mathbf{H}_{M \times Nt}$, where $M$ and $Nt$ are the number of antennas per user and per BS, respectively. In downlink precoding, we usually define the precoding unit (PU) for once precoding. And PU could be a resource block (RB), or many RBs, or across the whole bandwidth. If we set one PU includes $N_{RB}$ RBs, and one RB is with $N_{SC}$ marked subcarriers, there are $N_{RB}N_{SC}$ subcarriers, corresponding to the same granularity with SRS and $\mathbf{H}_{M \times Nt}$. That is to say that there are $N_{RB}N_{SC}$ SRS and $\mathbf{H}_{M \times Nt}$ in all for PU, and the channel reconstruction should be made from $N_{RB}N_{SC}$ MIMO channel $\mathbf{H}_{M \times Nt}$ to get one $\mathbf{H}_{S \times Nt}$ for once PU.

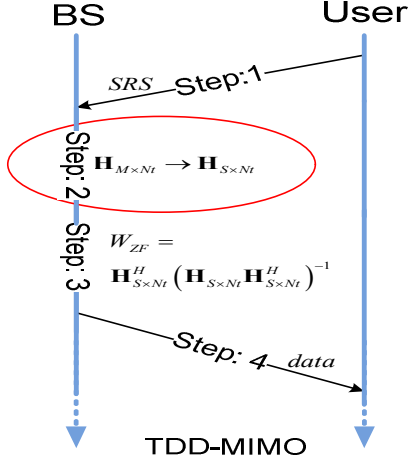

**Fig.3 Illustration of the downlink precoding process. For TDD systems, after the users' SRS in uplink, BS estimates SRS and gets the uplink channel $\mathbf{H}_{M \times Nt}$ corresponding to each user' $M$ antennas, and extracts effective channel $\mathbf{H}_{S \times Nt}$ for downlink precoding in step-2. Next is to process ZF precoding and to send data by $\mathbf{H}_{S \times Nt}$.**

The traditional method is eigen-beamforming (EBF), which calculates the correlation matrix and proceeds with the SVD, and here we call it *Direct SVD*. (Note that, here, SVD could be replaced by eigenvalue decomposition (EVD), as both of them hold the same process in nature and produce similar computation complexity.)

| *Direct SVD* |
|---|
| S-1: Calculate the average correlation matrix $\mathbf{R}$ based on $N_{RB}N_{SC}$ channels per PU. $$\mathbf{R}^k_{(Nt \times Nt)} = \frac{1}{N_{RB}N_{SC}} \sum_n^{N_{RB}N_{SC}} \left(\mathbf{H}^{n,k}_{(M \times Nt)}\right)^H \left(\mathbf{H}^{n,k}_{(M \times Nt)}\right)$$ |
| S-2: Get the main singular-vectors and reconstruct the channel. $$\left[\mathbf{v}^{1,k}_{(Nt \times 1)}, \mathbf{v}^{2,k}_{(Nt \times 1)}, \ldots, \mathbf{v}^{S,k}_{(Nt \times 1)}\right] = evd\left\{\mathbf{R}^k_{(Nt \times Nt)}\right\}$$ $$\left(\hat{\mathbf{H}}_k\right)_{S \times Nt} = \left(\mathbf{v}^{1,k}_{(Nt \times 1)}, \mathbf{v}^{2,k}_{(Nt \times 1)}, \ldots, \mathbf{v}^{S,k}_{(Nt \times 1)}\right)^H$$ |

In S-2, the core part is the EVD manipulation, which would produce large computation with $Nt$ increasing. In order to cut the computation, one potential solution is to reduce the matrix dimension in factorization, and another way is to design fast algorithms [4].

### 3.2 Proposed randomized method

Regarding to this problem, we all know that the EVD for $\mathbf{R}$ is equivalent to the SVD for $\mathbf{H}$. Here, we wish to construct an approximate SVD of a very large matrix $M \times Nt$ matrix $\mathbf{H}$. This can be achieved by getting randomized equivalent matrix and computing the SVD of the 'small' randomized matrix, as shown in Method I.

| *Method I* |
|---|
| S-1: Calculate the average channel matrix $\overline{\mathbf{H}}^k$ based on $N_{RB}N_{SC}$ channels per PU. $$\overline{\mathbf{H}}^k_{(Nt \times M)} = \frac{1}{N_{RB}N_{SC}} \sum_n^{N_{RB}N_{SC}} \left(\mathbf{H}^{n,k}_{(M \times Nt)}\right)^H$$ where $\mathbf{H}^{n,k}_{(M \times Nt)}$ is the channel matrix of the k-th user at the n-th subcarrier per PU. |
| S-2: Calculate the equivalent random matrix approximation Sample a random matrix $\mathbf{G}^k_{(M \times L)}$ with independent mean-zero, unit-variance Gaussian entries. $$\mathbf{Y}^k_{(Nt \times L)} = \overline{\mathbf{H}}^k \mathbf{G}^k$$ Construct $\mathbf{Q} \in \mathbb{R}^{Nt \times L}$ with columns forming an orthonormal basis for the range of $\mathbf{Y}$. For example, $\mathbf{Q}^k_{(Nt \times L)} = qr\left(\mathbf{Y}^k\right)$ |
| S-3: Calculate the SVD approximation Get the matrix $\mathbf{C}_{(L \times M)}$ $$\mathbf{C}_{(L \times M)} = \mathbf{Q}^H \overline{\mathbf{H}}^k$$ Calculate the singular-vector by SVD $$\mathbf{C} = \hat{\mathbf{U}}_{(L \times L)} \mathbf{\Sigma}_{(L \times M)} \mathbf{V}^H_{(M \times M)}$$ Get the effective channel $\mathbf{H}^k_{(S \times Nt)}$ $$\mathbf{H}^k_{(S \times Nt)} = \left(\mathbf{Q}_{(Nt \times L)} \widetilde{\mathbf{U}}_{(L \times S)}\right)^T$$ where $\widetilde{\mathbf{U}}$ is the $S$ vectors corresponding to the most $S$ eigen values in $\mathbf{\Sigma}$. and the approximate SVD of $\overline{\mathbf{H}}^k$ is displayed as $$\overline{\mathbf{H}}^k_{(Nt \times M)} \approx \mathbf{Q}_{(Nt \times L)} \hat{\mathbf{U}}_{(L \times L)} \mathbf{\Sigma}_{(L \times M)} \mathbf{V}^H_{(M \times M)}$$ |

The algorithm is simple to understand: $\mathbf{Y}^k_{(Nt \times L)} = \overline{\mathbf{H}}^k \mathbf{G}^k$ is an approximation of the range $\overline{\mathbf{H}}^k$; We therefore project the columns of $\overline{\mathbf{H}}^k$ onto this approximate range by means of the orthogonal projector $\mathbf{QQ}^H$ and hope that $\overline{\mathbf{H}}^k \approx \mathbf{QQ}^H \overline{\mathbf{H}}^k$. Next, in S-3, the point here of course is that the matrix $\mathbf{C}_{(L \times M)} = \mathbf{Q}^H \overline{\mathbf{H}}$ is $L \times M$ - we typically have $L \prec M \ll Nt$ - the computation cost of forming its SVD is on the order of $O(L^2 M)$ flops and fairly minimal.

Compared to Direct SVD, the method cuts the computation of the correlation matrix **R**, which reduces large amount of multiplication manipulation. Besides, the SVD in S-3 is only limited to $L$ and $M$, and the increasing of $Nt$ doesn't affect the complexity of the SVD. Although the QR factorization is used to construct **Q** in S-2, many simple implementation methods could replace QR factorization. So the complexity of constructing **Q** could be negligible.

This method requires $M \geq L \geq S$. And the selection of $L$ needs the balance between the performance and the complexity. On the one hand, $L$ is involved into the QR and SVD factorization and directly affects the complexity. So little $L$ saves much complexity. On the other hand, the approximation matrix $\mathbf{Y}_{(Nt \times L)}^{k}$ with larger $L$ could save more information in $\mathbf{H}_{(M \times Nt)}^{n,k}$. That is to say, larger $L$ could be with better performance.

Next, for this method, the natural interest is the accuracy of this procedure: how large is the size of the approximate residual? Specially, how large is $\left\| \overline{\mathbf{H}} - \mathbf{Q}\widehat{\mathbf{U}}\mathbf{\Sigma}\mathbf{V}^{H} \right\|_{F}^{2}$. This value could be proved by the following theorem.

**Theorem 1.** Let $\overline{\mathbf{H}}$ be an $Nt \times M$ matrix, $L = d + p$, $\mathbf{Q}^{H}\overline{\mathbf{H}} = \widehat{\mathbf{U}}\mathbf{\Sigma}\mathbf{V}^{H}$ and let $\sigma_i$ be the ith singular value of $\overline{\mathbf{H}}$, Then

$$\mathrm{E}\left\| \overline{\mathbf{H}} - \mathbf{Q}\widehat{\mathbf{U}}\mathbf{\Sigma}\mathbf{V}^{H} \right\|_{F}^{2} \leq \left(1 + \frac{d}{p-1}\right)\sum_{i>d}\sigma_i^2$$

*Proof:* See Appendix A

In the theorem, there is a limited approximation error in the Frobenius sense with different $d$ and $p$. If we set $d = 0$, the upper bound is the sum of singular value and the approximation error is heavily affected by the spectrum of the matrix $\overline{\mathbf{H}}$. If we set $d = L$, with $L$ increasing, then the upper bound is reduced and the gap becomes small. When $d = L = M$, we can get $\sum_{i>d}\sigma_i^2 = 0$, the proposed method approaching the Direct SVD. This ensures an acceptable performance in theory.

**3.3 Complexity analysis**

In this section, we analyze the computation complexity of the proposed method and compare it with the complexity of the classical SVD (used in Direct SVD). We express the computation complexity in terms of the number of floating point operations (FLOPs). In the following, each scalar complex addition or multiplication is counted as one FLOP. For the sake of simplicity, we do not distinguish between real-valued and complex-valued multiplications and neglect the computation complexity of common parts among these methods. Note that, although it cannot characterize the true computation complexity, FLOP counting captures the order of the computation load, so suffices for the purpose of the complexity analysis. Note that the exact number of calculation depends on the difference between implementation methods. Here, we take some popular calculation rules in [4] [10], which are widely used in methods analysis. Note that, considering that comparisons with other recent contributions or some fast methods for this model, which would provide a better justification and relative benefit of this proposed scheme, Jacobi algorithms may be a good choice in [4].

The following table gives the necessary calculation conclusion of typical processes. Here, $\mathbf{A} \in \mathbb{C}^{M \times N}$, $\mathbf{C} \in \mathbb{C}^{M \times N}$, $\mathbf{B} \in \mathbb{C}^{N \times L}$ are arbitrary matrices. $\mathbf{U}$, $\mathbf{\Sigma}$, $\mathbf{V}$ are the corresponding decomposition matrix of $[\mathbf{U}, \mathbf{\Sigma}, \mathbf{V}] = svd(\mathbf{A})$. $\mathbf{Q}$ is the orthogonal matrix of $\mathbf{A}$ by QR decomposition.

**Table I** *Float computation of traditional matrix model*

| | |
|---|---|
| Matrix-Matrix Prod of $\mathbf{A}\mathbf{A}^{H}$ | $M^2N + M(N - \frac{M}{2}) - \frac{M}{2}$ |
| Matrix-Matrix Prod of $\mathbf{A}\mathbf{B}$ | $2MNL - ML$ |
| Matrix Scaling of $\alpha\mathbf{A}$ | $MN$ |
| Matrix-Matrix Sum of $\mathbf{A} + \mathbf{C}$ | $MN$ |
| QR Matrix decomposition for $\mathbf{A}$, required $\mathbf{Q}$ | $4(M^2N - MN^2 + \frac{N^3}{3})$ |
| SVD Matrix decomposition for $\mathbf{A}$, required $\mathbf{\Sigma}, \mathbf{U}$ | $4M^2N + 13N^3$ [4] |
| SVD Matrix decomposition for $\mathbf{A}$, required $\mathbf{\Sigma}, \mathbf{V}$ | $2MN^2 + 13N^3$ [4] |

For both of the two methods, the matrix decomposition occupies most of the float computations. Meanwhile, the complexity order mainly depends on the dimension of the matrix, the times and the styles of decomposition method. Further, we give an accurate float computation results, including matrix decomposition, addition and multiplication. Note that, the common part of float computation and the manipulation beyond the discrimination in *Direct SVD*, *Method I* are ignored. Method I requires 1 SVD. And for constructing $\mathbf{Q}$, we consider QR factorization, which is thought as the upper bound of complexity in this step. In the following simulation, we set the $S = 2$, the first parameter of RO(*,*) is antennas of user $M$ and the second is the parameter $L$. The other parameters are based on Table II.

Fig.4 shows that, with the BS antennas $Nt$ increasing, both produce more float manipulations. Compared to Direct SVD, Method I only requires less than 10% float computations when the number of antennas is near 200 antennas with RO(8,2). Larger $Nt$ could enlarge the dual between Method I and Direct SVD.

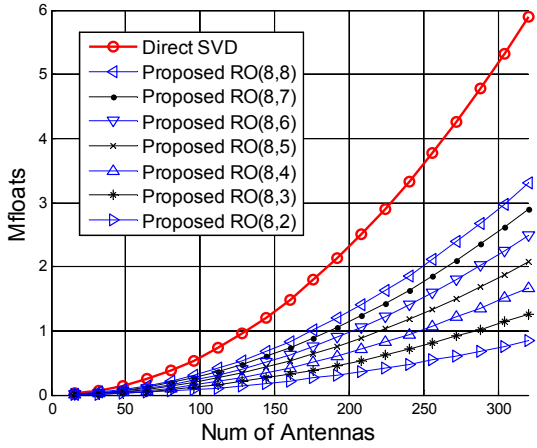

**Fig.4 Complexity comparison between proposed Method I and Direct SVD, and the simulation parameters are based on the following Table II.**

In Method I, the SVD is on basis of $M$ and $L$, so with $L$ increasing, RO (*,*) results in different complexities. Larger $L$ causes more computation. What's more, the construction in Q is with $Nt$ and $L$. So with Nt increasing, the complexity also increases.

## IV. SIMULATION RESULTS

In this section, we provide some numerical results to evaluate our proposed low-complexity channel reconstruction methods. The system model considered is in Section II with a BS employing a $8\times16$ UPA with 7 users, each having 8 antennas. The 3D-MIMO-UMi channels are modeled as [9]. Table II lists the detailed simulation parameters.

**Table II** *3D-MIMO simulation parameters*

| Parameters | Settings |
|---|---|
| Scenario | 3D-MIMO-UMi |
| User antenna configuration | 8Rx, dual-polarized (0/+90) |
| BS antenna configuration ($H\times V$) | Dual-polarized $8H\times 8V$ (128 antennas) |
| Bandwidth | 20MHz (100 RBs) |
| Antenna element interval | 0.5 carrier wave length both in horizontal and vertical direction |
| Carrier frequency | 2GHz |
| Number of users | 7 |
| Number of streams per user | 2 |
| MCS | Fixed, 64QAM |
| User distribution | Referred to [11] |
| User speed | 3km/h |
| Traffic model | Full buffer |
| Channel estimation | Ideal |
| Receiver | MMSE-IRC |

Considering the link simulation with MU-MIMO systems, Fig.5 exhibits the performance of the proposed methods with different random matrix dimensions ($L=2,4,6,8$) in $M=8$, $S=2$, $Nt=128$. When SNR=40dB and $L\geq 4$ in RO(8,$L$), Method I could achieve beyond 1Gbps. $L=8$ in RO(8,$L$) could totally match the Direct SVD at SNR>10dB. Note that, with SNR increasing, the gap is enlarging between Direct SVD and Method I with less $L$, e.g., $L=2$ in RO(8,$L$).

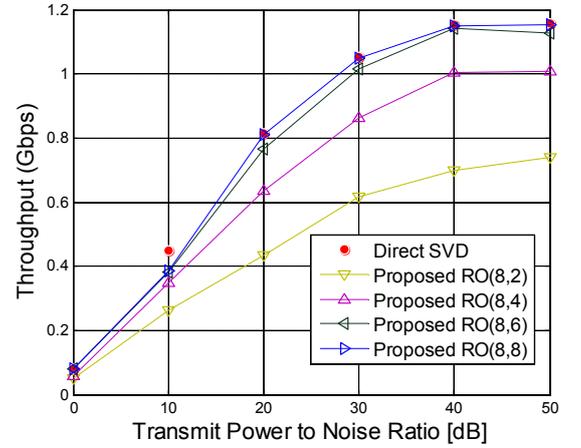

**Fig.5 Achievable throughput using the proposed method compared with the ideal Direct SVD, with different randomized matrix dimensions.**

If we consider the gap between Method I and Direct SVD in Fig.6, the metric is defined as RO/Direct SVD (%), which displaces the percent over Direct SVD. When $L\geq 6$ in RO(8,$L$), it displays excellent performance and approaches to the ideal Direct SVD. And $L=8$ gets the best performance.

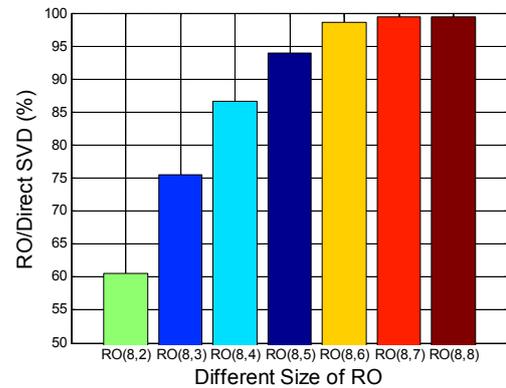

**Fig.6 Achievable throughput using Method I, with different random matrix dimension RO(M,L) on SNR=40dB.**

Based on the analysis and simulation results, we can conclude that the complexity and throughput are opposed to each other. Less $L$ leads to less float computation, but the throughput also gets less. In future 5G, while meeting the performance requirements, a reasonable complexity is necessary. And referring to the simulation, the random matrix dimension needs to be revised. From the above two figures, this parameter optimization is necessary with the demands of the performance and the computation complexity. The simulation results match our traditional expectation: the higher complexity method gets better performance. And the simulation is based on realistic parameters in 3D-MIMO, and the results could be considered as a valuable performance target.

## V. CONCLUSIONS

This paper proposes a series of low-complexity channel reconstruction methods for TDD massive MIMO systems. These methods fully explore the characters of low-rank channel matrix from BS to each user, and transform the large redundant range into small approximating range by sampling a random matrix. Compared to the existing method, they could largely reduce the computation complexity, especially with large-scale antennas. The complexity analysis shows that the proposed methods only require less than 30% float computations, while the performance could kept around 1Gbps by downlink ZF precoding.

## ACKNOWLEDGEMENT

This work was supported by the National High Technology Research and Development Program of China (863 Program) (Grant No. 2014AA01A705).

## References
[1] Marzetta, Thomas L. "Noncooperative cellular wireless with unlimited numbers of base station antennas." *Wireless Communications, IEEE Transactions on* 9.11 (2010): 3590-3600.
[2] Alcatel Lucent, Lightradio, http://www.alcatel-lucent.com/lightradio/
[3] Nokia Simense Networks, Liquid radio, http://www.nokiasiemensnetworks.com/portfolio/liquidnet/liquidradio/.
[4] Golub, Gene H., and Charles F. Van Loan. *Matrix computations*. Vol. 3. JHU Press, 2012.
[5] Saad, Yousef. *Iterative methods for sparse linear systems*. Siam, 2003.
[6] Abed-Meraim, K., A. Chkeif, and Y. Hua. "Fast orthonormal PAST algorithm." *Signal Processing Letters, IEEE* 7.3 (2000): 60-62.
[7] Ren, Yuwei, et al. "Low-Complexity Channel Reconstruction Methods Based on SVD-ZF Precoding in Massive 3D-MIMO Systems." http://arxiv.org/abs/1510.08507
[8] Li, Yan, et al. "Dynamic beamforming for three-dimensional MIMO technique in LTE-advanced networks." *International Journal of Antennas and Propagation* 2013 (2013).
[9] 3GPP,TR, 36.873, v1.3.0 (2014-02) Study on 3D channel model for LTE(*Release 12*).
[10] Hunger, Raphael. *Floating point operations in matrix-vector calculus*. Munich: Munich University of Technology, Inst. for Circuit Theory and Signal Processing, 2005.
[11] Lee, Chang-Shen, et al. "Sectorization with beam pattern design using 3D beamforming techniques." *Signal and Information Processing Association Annual Summit and Conference (APSIPA), 2013 Asia-Pacific*. IEEE, 2013.
[12] Witten, Rafi, and Emmanuel Candès. "Randomized algorithms for low-rank matrix factorizations: sharp performance bounds." *Algorithmica* 72.1 (2013): 264-281.

# Appendix

## A. Proof for Theorem 1

Before the proof, let $\mathbf{D}_{M-d}$ be the diagonal matrix of dimension $M-d$ equal to $diag(\sigma_{d+1}, \sigma_{d+2}, \ldots, \sigma_M)$ and define $f(\overline{\mathbf{H}}, \mathbf{G}) := (\mathbf{I} - \mathbf{Q}\mathbf{Q}^H)\overline{\mathbf{H}}$. So,

$$\left\|\overline{\mathbf{H}} - \mathbf{Q}\widehat{\mathbf{U}}\boldsymbol{\Sigma}\mathbf{V}^H\right\|_F^2 \overset{a}{=} \left\|\overline{\mathbf{H}} - \mathbf{Q}\mathbf{Q}^H\overline{\mathbf{H}}\right\|_F^2$$
$$= \left\|(\mathbf{I} - \mathbf{Q}\mathbf{Q}^H)\overline{\mathbf{H}}\right\|_F^2 = \left\|f(\overline{\mathbf{H}}, \mathbf{G})\right\|_F^2$$

The equation 'a' is from $\mathbf{Q}^H\overline{\mathbf{H}} = \widehat{\mathbf{U}}\boldsymbol{\Sigma}\mathbf{V}^H$. In [12, Theorem 1.4], the worst case error for matrices with such singular values is equal to the random variable

$$W(\mathbf{D}_{M-d}) = \left\|f(\mathbf{D}_{M-d}, \mathbf{X}_2)\left[\mathbf{X}_1\boldsymbol{\Sigma}^{-1} \; \mathbf{I}_{M-d}\right]\right\|_F^2$$

Here, $\mathbf{X}_1$ and $\mathbf{X}_2$ are respectively $(M-d) \times d$ and $(M-d) \times p$ matrices with i.i.d. $CN(0,1)$ entries, $\boldsymbol{\Sigma}$ is a $d \times d$ diagonal matrix with the singular values of a $(p+d) \times d$ Gaussian matrix with i.i.d. $CN(0,1)$ entries, and $\mathbf{I}_{M-d}$ is the identity matrix. Furthermore, $\mathbf{X}_1, \mathbf{X}_2$ and $\boldsymbol{\Sigma}$ are all independent (and independent from $\mathbf{G}$).

So,

$$\left\|f(\overline{\mathbf{H}}, \mathbf{G})\right\|_F^2 \leq \left\|f(\mathbf{D}_{M-d}, \mathbf{X}_2)\left[\mathbf{X}_1\boldsymbol{\Sigma}^{-1} \; \mathbf{I}_{M-d}\right]\right\|_F^2$$
$$\overset{b}{\leq} \left\|\mathbf{D}_{M-d}\left[\mathbf{X}_1\boldsymbol{\Sigma}^{-1} \; \mathbf{I}_{M-d}\right]\right\|_F^2 \overset{c}{=} \left\|\mathbf{D}_{M-d}\mathbf{X}_1\boldsymbol{\Sigma}^{-1}\right\|_F^2 + \left\|\mathbf{D}_{M-d}\right\|_F^2$$

The inequality 'b' follows from the fact for any orthogonal projector $\mathbf{P}$, $\left\|\mathbf{P}\overline{\mathbf{H}}\right\|_F^2 \leq \left\|\overline{\mathbf{H}}\right\|_F^2$, and the equation 'c' is Pythagoras' identify. Now a simple calculation we omit gives

$$\mathrm{E}\left\|\mathbf{D}_{M-d}\mathbf{X}_1\boldsymbol{\Sigma}^{-1}\right\|_F^2 = \left\|\mathbf{D}_{M-d}\right\|_F^2 \mathrm{E}\left\|\boldsymbol{\Sigma}^{-1}\right\|_F^2$$

So we can get the final expression by $\left\|\mathbf{D}_{M-d}\right\|_F^2 = \sum_{i>d}\sigma_i^2$ and $\mathrm{E}\left\|\boldsymbol{\Sigma}^{-1}\right\|_F^2 = d/(p-1)$ [12, Corollary 1.3].